# Geometrical Tailoring of Shockley-Ramo Bipolar Photocurrent in Self-powered GaAs Nanodevices

Xiaoguo Fang[1,2,#], Huanyi Xue[2,3,4,#,*], Xuhui Mao[3], Feilin Chen[2], Ludi Qin[3], Haiyue Pei[2,5], Zhong Chen[6], Pingping Chen[7,8], Ding Zhao[2,4,5], Zhenghua An[3,*], Min Qiu[2,4,5,*]

1. College of Information Science and Electronic Engineering, Zhejiang University, Hangzhou, Zhejiang 310027, China.
2. Zhejiang Key Laboratory of 3D Micro/Nano Fabrication and Characterization, Department of Electronic and Information Engineering, School of Engineering, Westlake University, Hangzhou, Zhejiang 310030, China.
3. State Key Laboratory of Surface Physics and Institute for Nanoelectronic Devices and Quantum Computing, Department of Physics, Fudan University, Shanghai, 200438, China.
4. Westlake Institute for Optoelectronics, Fuyang, Hangzhou, Zhejiang 311421, China.
5. Institute of Advanced Technology, Westlake Institute for Advanced Study, 18 Shilongshan Road, Hangzhou 310024, Zhejiang Province, China.
6. Instrumentation and Service Center for Molecular Sciences, Westlake University, 600 Dunyu Road, Hangzhou 310030, Zhejiang Province, China.
7. State Key Laboratory of Infrared Physics, Shanghai Institute of Technical Physics, Chinese Academy of Sciences, Shanghai 200083, China.
8. University of Chinese Academy of Sciences Beijing 101408, China.

[#] These authors contributed equally to this work.
**\* Authors to whom correspondence should be addressed:** xuehuanyi@wioe.westlake.edu.cn; anzhenghua@fudan.edu.cn; qiu_lab@westlake.edu.cn

## Abstract

Bipolar photoresponse—where photocurrent polarity reverses with excitation wavelength, gate voltage, or other conditions—is essential for optical logic, neuromorphic computing, and imaging. Unlike unipolar responses, bipolar behavior enables direct binary encoding and enhanced photodetection contrast. However, in conventional photoconductive or photovoltaic systems, the simultaneous and opposite-directional transport of electrons and holes often suppresses polarity switching. Recent self-powered Shockley-Ramo (SR) photoresponse in gapless materials also show only unipolar signals due to strong, irreversible electron-hole asymmetry. Here, we demonstrate for the first-time bipolar SR photoresponse in GaAs nanoconstriction devices by exploiting reversible electron-hole asymmetry. The longer carrier lifetimes in GaAs enable sub-diffusion-length control of carrier dynamics through geometry. By tuning photocarrier dynamics near the nanoconstriction for both majority electrons and minority holes, we modulate the SR response to exhibit dual polarities. At low excitation, photoelectrons dominate; as excitation increases, intervalley scattering populates higher-energy *L*-valleys, reducing electron contribution and leading to polarity reversal driven by the growing dominance of photoexcited holes. These results, supported by SR theory, show that nanoscale geometric engineering, together with the reversible electron-hole asymmetry, enables self-powered bipolar photocurrent responses, offering new routes toward advanced optoelectronic devices.

# 1. Introduction

Transforming photon energy into charge motion is a fundamental light–matter interaction process underlying all optoelectronic devices, and the resulting measurable photoresponse is essential for applications ranging from photovoltaic solar cells to photodetectors[1-3]. A deep understanding of the underlying mechanisms is crucial for enhancing optical-to-electrical conversion efficiency[4,5], improving device performance[6], and innovating new functionalities such as optically encoding[7]. For the binary encoding applications, self-powered bipolar photoresponse offer a straightforward and superior means to encode binary digital information and are hence desirable from the application point-of-view[8-13]. To realize bipolar photoresponse, however, conventional photodetectors based on photoconductive or photovoltaic mechanisms face challenges in producing bipolarity because, they rely on either an externally applied or internally built-in electric field and, electrons and holes typically exhibit opposite drift directions and contribute to the final photocurrent constructively due to their opposite charges, leaving the realization of single-wavelength bipolar response difficult[8-13]. Besides, the short-lived photo-charges typically transport a limited length being substantially shorted than the distance between macroscopic electrodes[13,14]. Consequently, each absorbed photon often produces less than one measurable photo-charge in external circuits unless some other mechanisms are introduced such as photogating[15,16] to dramatically increase one type of photocarrier lifetime and hence the photoconductive gain, resulting in unipolar photoresponse due to the selective prolonged lifetime of the one-type photocarrier.

Recently, an alternative mechanism utilizing geometrical low-symmetry edges or corners have been shown to induce the global photoresponse via the so-called Shockley-Ramo (SR) approach[17]. This provides a promising solution to address the above two challenges simultaneously. First, the instantaneous transport directions of photo-charges are defined by the local geometrical asymmetry and are the same for electrons and holes, unlike their transport in electric fields. This allows one to determine the polarity by selectively choosing which carrier leads to the final photo response. Second, in this SR scheme, the self-powered (zero-bias) long-range photoresponses are generated by the instantaneous electric currents from moving photocharges, despite of their short lifetimes, because they arise from the self-induced electric field flux perceived by each electrode, rather than the net amount of photocharge entering the electrode per second[17-19]. Consequently, the induced currents are only weakly sensitive to the charge position but depend strongly on the dominant charge type and the charge velocity—both its magnitude and direction. To date, SR-type photoresponses have been reported mainly in two-dimensional gapless materials such as graphene[20,21], Weyl semimetals ($MoTe_2$, $WTe_2$, and $TaIrTe_4$)[22-24], and topological insulators ($Bi_2Se_3$ and $BiSbTe_3$) [25,26]. Despite the diverse physics driving the photoresponse in these gapless materials, the SR mechanism plays a commonly important role in their global photoresponse. On the other hand, these reported works utilize only the majority carriers and hence realize only one polarity due to the strong but irreversible electron-hole asymmetry.

Here, we demonstrate for the first time SR-type photoresponses for gapped materials, to the best of our knowledge. It is generally assumed that slow recombination of photo-excited carriers in gapped materials readily induces spatial nonlocality in the optoelectronic response. As a result, little attention has been paid thus far to the role of SR mechanisms in their global photoresponse. We geometrically tailor the global photocurrent by utilizing the asymmetric band structures for majority electrons and minority holes, controlling their populations in multiple energy valleys of GaAs. As a classical semiconductor, GaAs has been widespread in modern optoelectronics, yet the microscopic nanodynamics of GaAs remains an intriguing field, with emergent exotic properties related to nanoscale nonequilibrium hot electrons reported in recent years. These include nonlocal energy dissipation[27], phonon-bottlenecked quasi-adiabatic transport[28], and anisotropic hot-electron domains[29]. The tailorable competitive and imbalance of electron and hole transport in the sub-diffusion-length-scale present in this work opens a new pathway to further programming and optimizing the performance and functionalities of current optoelectronics.

## 2. Results and Discussions

In our experiments, the GaAs-based devices consist of quasi-two-dimensional electron gas (q-2DEG) layer in (001) face of *n*-doped GaAs quantum well (QW) with 35-nm-thickness that confined by $Al_xGa_{1-x}As$ barrier layers (Figure S1, Supporting Information). The geometrically patterned conducting channels are fabricated by a standard electron beam lithography (EBL) and two-step etching process that connected to the drain/source electrodes via Ohmic contacts (see Experimental Section and Figure S2, Supporting Information). The distance between drain and source electrodes is fixed to be 170 μm for all samples, much larger than the characteristic lengths of charge transport such as mean free path (~55 nm)[27], energy relaxation lengths (~200–300 nm)[27] and diffusion lengths (~several microns)[30]. Note that the channel region between the two electrodes is composed solely of a single GaAs QW material, with no lateral hetero- or homo-junctions present to induce a conventional photoresponse. Details for the sample information and device fabrication are same as in reference[27-29]. Scanning photocurrent microscopy (SPCM) measurements are implemented by scanning a focused laser spot (532 nm continuous wave, 2.33eV) over the device and recording the short-circuit current through the drain/source electrodes at zero bias voltage, as schematically shown in Figure 1a. The photon energy of laser beam is sufficiently high to allow population of high energy valleys when the excitation power is large enough[31,32].

We first investigated the photocurrent (PC) response in a pristine quasi-two-dimensional electron gas (q-2DEG) channel (width = 40 μm) without nanostructures. As shown in Figure 1b, the PC response (up to ~100 nA) predominantly localized at the channel edges perpendicular to the channel axis (oriented along the x-direction), exhibiting negative polarity at the left edge and positive polarity at the right edge, respectively. By fabricating a nano-constriction structure (width = 600 nm) on the center of the pristine q-2DEG channel, additional PC signals with enhanced intensity (~260 nA) were generated around the structure. These signals anti-symmetrically distributed on both sides of constriction with opposite polarities, suggesting the effectiveness of local patterning in generating global photoresponse. This PC signal increases monotonously as the width of the constriction structure shrinks from 20 μm to 0.6 μm, (Figure 1d).

Notably, for the narrowest case (0.6 μm), PC signals with the same sign extend from edges toward the interior region and merge together, clearly demonstrating a pronounced enhancement of the PC response by constriction. This behavior highlights the significant influence of geometric confinement on the delocalization and intensification of PC signals. In addition, we find that these PC signals remain stably concentrated around the constriction regardless of its relative position within the pristine channel (see Figure S3, Supporting Information). These zero-bias PC responses, characterized by geometry-sensitive and long-range response features, cannot be explained by the conventional drift-diffusion model[33-35], which solely relies on limited photocarrier diffusion lengths and relaxation times to describe the amount of charge reaching the electrodes per second, because: (1) no externally applied bias to drive the drift motion of photocarriers; (2) the distance between electrodes are far larger than the diffusion lengths of photocarriers. Instead, they can be well described by the SR-type approach (as illustrated in Figure 1e): the PC responses originate from two symmetry-breaking mechanisms[17-26]: (1) the *e-h* asymmetry is a prerequisite for generating the local injection current within the excitation region. Under the quasi-equilibrium approximation (as will discussed later), this local current is predominantly governed by diffusion of photoelectrons, as the room-temperature (300 K) diffusion ability $D$ of electrons is significantly higher than that of holes that determined by the electric mobilities $\mu_{e/h}$ of electrons/holes in our GaAs q-2DEG according to the Nernst-Einstein relation[36]:

$$D_{e/h} = \mu_{e/h} k_B T / q \tag{1}$$

; (2) the spatial circular asymmetry ensures an anisotropic propagating mode, yielding a net local current $\boldsymbol{j}_{local}$ in the low-symmetry regions, such as edges and entrances or exits of constrictions. This net current, in turn, determines the polarity of the photocurrent (PC) according to the orientation of the structure. Finally, $\boldsymbol{j}_{local}$ generates a long-range PC response in accordance with the SR theorem[17]:

$$I_{ph} \propto \int \boldsymbol{j}_{local} \cdot \nabla \psi(\mathbf{r}) d^2 \mathbf{r} \tag{2}$$

where $\psi$ is the auxiliary weighing field. To verify this assumption, we performed quantitative simulations and measurements by extracting the angle-dependent PC response along the arc edges of the constricted geometry. The simulated results show excellent agreement with experimental data (see Note S1 and Figure S4, Supporting Information), providing preliminary evidence for the SR-type photoresponse shown in Figure 1d.

Next, we systematically investigated the excitation power dependence of the PC response. Figure 2a presents large-scale SPCM images of the entire channel at laser powers ranging from 3 to 11 mW for the sample with the narrowest width (0.6 μm). The overall PC signals display similar characteristics with same polarities and sub-linearly increasing intensities across the tested power range. A more detailed analysis of high-precision line profiles (along the channel, Figure 2b) and local-scale maps (insets of Figure 2b) reveals that an additional narrow PC feature emerges in the close vicinity of the constriction. As the laser power increases, these features grow remarkably with the polarity reversed relative to the original PC response. The amplitude of this abnormal signal can approach the maximum of the original signals (bottom panel of Figure 2b). The emergence of this power-dependence bipolar signal implies that opposite charge carriers, holes, may contribute to this photoresponse. This is, however, very surprising because QW is electron-doped and holes remain the minority carriers since photoexcitation creates electrons and holes always in pairs, irrespective of the excitation power, and the photo-generated electrons/holes are approximately 20 percent of the ambient electron density under 1-mW laser excitation (Figure S2c, Supporting Information). Power-dependent bipolar PC responses were further measured at several fixed positions along the edge of the constriction in close proximity to the narrowest region (marked by colored dots in the inset of Figure 2c). As depicted in Figure 2c, the PC response distinctly demonstrates a non-monotonic behavior with increasing power. Initially, the signals exhibit a nonlinear increase, approaching saturation, and are followed by a steady decline with a tendency to reverse sign as the laser power significantly exceeds the threshold for saturation PC. The non-monotonic characteristic is sensitive with local geometry that becomes more prominent as the irradiation position approaches the narrowest region, as demonstrated by the shift of the saturation power (highlighted by the green arrows in Figure 2c) and the threshold power to reverse the polarity. This behavior is in stark contrast to the conventional photoelectric phenomena, where the PC response typically exhibits a monotonic dependence on power, indicating a complex interplay between geometric confinement and excitation dynamics in our bipolar PC response. Additionally, we checked that the characteristic bipolarity was absent in broader-constriction samples (8 μm in width) even excited under higher excitation power (Figures S5 and S9, Supporting Information), thereby providing further support for the intrinsic correlation between geometrical confinement and anomalous bipolar characteristics in Figure 2.

To understand the physical mechanism behind the observed bipolar PC response—where both majority carriers (electrons) and minority carriers (holes) contribute—we must consider the local dynamics and transport of photo-created carriers in the close vicinity of nanoconstriction channel upon photoexcitation, until their recombination and annihilation. The relaxation dynamics of photo-excited hot carriers is sketched in Figure 3. Under the laser excitation, electrons in the valence band are photoexcited into the Γ valley of the conduction band, generating hot electrons and leaving holes (mainly heavy holes, HHs[37]) in the valence band. These hot electrons would dissipate their excess energy to the surrounding electrons gas via *e-e* interaction, namely, electron thermalization (process ①) or release to the lattice through electron-phonon (*e-ph*) interaction (process ②)[27-29,38]. Simultaneously, the high-energy tail of the hot Γ-electrons may scatter into the higher-energy *L* valley (process ③)[39,40]. Under low-power excitation conditions, the contribution from process ③ is not prominent because *L*-electrons can rapidly back-scatter into Γ-valley via emission of phonons with large wave vector ($q$) and then contribute to photoresponse in the same manner as Γ-electrons. Hence, it can be regarded that, at low excitation conditions, the steady-state population of photoelectron within Γ valley ($n_{e,\Gamma}$) approximatively equal to that of HHs ($n_h$), as sketched in left side of Figure 3b. This leads to the dominance of photoelectrons to the observed PC response because HHs have much larger effective mass and lower mobility[35]. Under high-power excitation, however, electron population (process ③) in *L*-valleys increases rapidly and becomes prominent, reminiscent of the well-known high-electric-field induced Gunn effect[41]. Unlike the electric-field-induced Gunn effect case, the appreciable population of *L*-valley here is enabled by the high optical pumping intensity because significant portion of Γ-electrons (including both non-thermalized energetic electrons with large partitioned energy from pumping photons and the high-energy-tailing distribution of thermalized hot Γ-electrons) have sufficiently high energy ($E \geq E_L$) to allow forward intervalley scattering of electrons (Γ→*L*) by emission or absorption of large-$q$ LO phonons, until it is balanced by the increasing back intervalley scattering (Γ←*L*). When this balanced steady-state condition is reached (Γ↔*L*), additional pumping will further increase mostly the population of *L*-electrons, i.e., $n_{e,L}$.

In this model, as illustrated in Figure 3c,d, the dashed blue/red circles represent the diffusion boundaries of *e*/*h* carriers injected from the circle centers (laser excitation position), with radii corresponding their respective diffusion lengths[36]

$$L_{D,i} = \sqrt{D_i \tau} \tag{3}$$

where $\tau$ denotes the minority carriers (holes) lifetime and $D_i$ corresponds to the diffusion coefficient of *e*/*h* carriers (superscript *i* denotes *e* or *h*), which is proportional to mobility according to Eq. (1). The diffusion lengths of Γ-electrons $L_{D,e}$ and holes $L_{D,h}$ at 300 K were estimated to be approximately 10 and 4 μm, respectively, using Eq. (3) (see Note 2, Supporting Information). Geometrically, the constriction structures impose the geometrical confinement on the diffusion of

$e/h$ carriers by truncating their diffusion path. However, the large difference between $L_{D,e}$ and $L_{D,h}$ leads to the distinct responses in confinement effect for electrons and holes. When the injection position is far from the constriction (denoted as region I in Figure 3c,d), geometry boundaries exert minimal influence on carrier diffusion paths, resulting in zero local current due to isotropic diffusion[20]. As the position approaches the constriction (region II), electron diffusion boundaries were truncated by constriction edges while the diffusion of holes remains almost isotropic, generating net $j_e$ and weak $j_h$ ($j_e > j_h$), thus producing a negative local current:

$$j_{local} = -j_e + j_h \qquad (4)$$

on left side of constriction. In this regime, PC signal intensity and polarity are primarily governed by electron diffusion (Γ-valley) with superior mobility and longer diffusion length, aligning with the physical mechanism shown in Figure 1e. The scenario becomes more complex when injection occurs in the vicinity of the constriction (region III), where both $e/h$ carrier diffusion paths are truncated. In this configuration, the constriction width $W$ plays a significant role in modulating the $j_{local}$ due to the fact that the $j_h$ exhibits higher sensitivity to the variation of $W$, for simplicity considering the low excitation condition ($n_{e,\Gamma} \approx n_h$), as discussed below: As shown in Figure 3c, if $W$ is comparable or larger to $L_{D,h}$ (more exactly, $2L_{D,h}$), the constriction is less effective in confining hole diffusion while remaining capable for restricting electron diffusion, hence in this regime, $j_e \gg j_h$ (Figure 3e), indicating $j_{local}$ is governed by $j_e$. In contrast, if $W$ is shrinking to a narrow scale, $W \ll L_{D,h}$, the diffusion of holes is significantly confined by the constriction, causing a considerable $j_h$ compared to the $j_e$ in this regime as depicted in Figure 3f. As previously discussed, high excitation would reduce the ratio of $n_{e,\Gamma}/n_h$, therefore leading to a suppression of $j_e$ (relative to $j_h$) as indicated by the light blue lines in Figure 3e and 3f. Remarkably, this would induce a sign change of $j_{local}$ in close proximity to the constriction (region III) where $j_e < j_h$ (as indicated by the green arrow in Figure 3f), this bipolar response is more likely to be observed especially in the narrow-constricted device owing to the comparable intensities of $j_e$ and $j_h$ in region III (Figure 3f). With the power increase, the decrease of $n_{e,\Gamma}/n_h$ further enhances the contrast of abnormal signals, consistent with the experimental results (Figure 2) and numerical simulations (Note S3 and Figure S8, Supporting Information). Moreover, by shrinking constriction width the auxiliary weighing field $\nabla \psi$ could increase sharply in region III (as sketched in the bottom panels in Figure 3e,f), the abnormal PC signals are therefore significantly enhanced according to Eq. (2). This semi-quantitative analysis was validated for reproducing the power-dependent PC behaviors in numerical simulations, which exhibited excellent agreement with the experimental data (Figure S8, Supporting Information). Notably, shrinking constriction width not only enhances the confinement of hole diffusion, but also significantly sharpens the $\nabla \psi$ near the constriction (Note S3 and Figure S9, Supporting Information)), thereby facilitate the anomalous PC signals.

As previously discussed, power-dependent intervalley scattering serves as a unique mechanism for modulating the ratio of $n_{e,\Gamma}/n_h$, thereby manipulating the bipolar PC responses with geometrical confinement. Alternatively, we noted that the asymmetric temperature-dependent mobilities of electrons and holes can also be utilized to modulate the bipolar photoresponse. At lower temperatures, the mobility of minority carriers (holes) in GaAs increases proportional to $\sim T^{-\alpha}$ ($\alpha \geq 1$) with decreasing temperature, whereas the mobility of majority carriers (Γ-electrons) remains nearly constant (see Note S2, Supporting Information). As mentioned earlier, the net diffusion current of electrons and holes ($j_{e/h}$) around the constriction region are primarily dominated by their diffusion lengths. This indicates that temperature reduction would also play an effective role for tailoring the global photoresponse by modulating the ratio of $L_{D,e}/L_{D,h}$, i.e., $\mu_e/\mu_h$, according to Eqs. (1) and (3). To verify, additional experiments were performed under a fixed high power (6 mW) across a low-temperature range (120 – 300 K), by systematically investigating the temperature-dependent photoresponse of 0.6-μm-constriction device. Figure 4a illustrates the large-scale PC distributions of a 0.6-μm-wide constriction device. Two distinct characteristics emerge with decreasing temperature: (1) the original PC signals (exhibiting features identical to those in Figure 1b), which dominate the constriction region II at 300 K, gradually diminish in intensity; (2) conversely, the anomalous PC signals, initially localized near the constriction center (region III) with slightly inversed intensity at 300 K (consistent with Figure 2a), progressively expand outward and ultimately occupy the entire constriction region at 120 K, accompanied by an almost complete polarity reversal across the area.

To quantitatively examine these evolving features, line-scanning measurements were performed along the entire channel (as indicated by the green dashed line in Figure 4a). As depicted in Figure 4b, the original PC signals display a gradual decay from ~600 nA (300 K) to nearly undetectable levels (less than 100 nA at 120 K) as temperature dropping. In contrast, the peak intensities of anomalous PC signals exhibit a monotonic increase from ~200 nA to ~600 nA, while their spatial widths broaden continuously with decreasing temperature. These results provide additional evidence for an unconventional PC response induced by geometrical confinement that exhibits the abnormal temperature-dependent evolutions in the PC behaviors. To elucidate these two opposite temperature-dependent trends in PC evolution at regions II and III, we performed the numerical simulations based on our model by taking into account the temperature dependent carriers mobilities $\mu_{e/h}(T)$ and the life time $\tau(T)$ in Eq. (3) (Note S2 and Figure S10, Supporting Information). Simulation data demonstrate that during the cooling process, $L_{D,e}$ decreases dramatically from ~10 μm (300 K) to 4 μm (120 K), whereas $L_{D,h}$ maintains nearly constant (see Note S2 and Figure S10, Supporting Information). This would diminish both the spatial range and peak magnitude of $j_e$ distribution profile, correspondingly suppress the intensity of $j_e$ around constriction regions. The PC signals in region II are predominated by $j_e$, where $j_e \gg j_h$ at 300 K (as illustrated in Figure 3), as temperature decrease the sustained decrease of $j_e$ gradually approaches $j_h$ at 120 K (inset of Figure 4c) significantly reduce the PC signals (as shown in Figure 4c) according to Eqs. (1) and (4). Furthermore, under the excitation power of 6 mW, the intervalley-scattering mechanism provides an initial suppression of $j_e$ at 300 K (illustrated in Figure 3f), leading to a slight overlap between $j_e$ and $j_h$ in region III (where, $j_e < j_h$, as illustrated in the inset of Figure 4d) with the reversed-polarities PC signals. Notably, as the intensity of $j_e$ is further suppressed under cryogenic temperature (inset of Figure 4d), the abnormal PC signals are therefore increased due to the enhanced overlap between $j_e$ and $j_h$ (Figure S11a, Supporting Information), leading to the opposite trend compared to that in region II. Both the temperature-dependent evolutions and line shapes of simulated PC line profiles (Figures S11 b-d, Supporting Information) are in excellent agreement with the experimental data, thereby further validating the proposed model based on photocarrier diffusion within the SR-type scheme.

The excellent agreement between our experimental observations and the simulated bipolar responses indicates that the global photoresponse is still predominantly governed by the local diffusion current driven by the carrier concentration gradient, $j_{local}(r) \propto \nabla n(r)$ [19], for both electrons and holes. This is consistent with our use of the carriers' diffusion constants and lifetimes in the simulation, instead of using thermal velocities and electron-phonon scattering times. It also implies that the contribution from the local thermoelectric current, $j_{local}(r) \propto \nabla T(r)$[19], arising from carrier temperature gradients, is negligible. This can be attributed to the fact that the energy relaxation time (~75 fs)[27] of photo-excited hot carriers—governed by electron-phonon scattering—is much shorter than their lifetime (~20 ns)[42], which is determined by recombination processes. Therefore, one might conclude that hot electron effects play a minor role in the observed global photoresponse. However, the observed polarity reversal at high excitation power reveals that hot electrons populated in the *L*-valley become effectively frozen and contribute little to the photoresponse. This suggests that, under high excitation, the intervalley backscattering from *L* to Γ is suppressed by the hot phonon bottleneck effect, preventing these carriers from relaxing to the conduction band minimum and participating in the photoresponse. Therefore, the bipolarity of the photoresponse observed in this work can be regarded as a manifestation of this phonon bottleneck effect in our high optical excitation condition, in contrast to the double electron temperature peak feature observed previously in high electric field case[28]. Finally, we note that, unlike in conventional photodetectors where background doping typically increases dark current and degrades device performance, the high ambient electron concentration in our system plays an essential role in mediating and enabling the SR-type photoresponse for both photoexcited electrons and holes. This ambient carrier population[17] is therefore crucial to the observed bipolar photoresponse in our study. These findings offer valuable insights for optimizing the performance of current optoelectronic devices and for exploring new functional modalities.

## 3. Conclusion

In summary, this work experimentally demonstrates that the nanoconstriction architecture serves as a versatile platform for investigating the complex interplay between geometric confinement, photo-charge dynamics, and intervalley scattering by using the spatially resolved PC measurements. The observed power- and geometry- dependence in bipolar PC response highlights the rich physics arising from non-equilibrium carrier dynamics in confined geometrical structures. In addition, these findings are further verified in the additional cryogenic experiments by controlling the ratio of $L_{D,e}/L_{D,h}$ for achieving the contrary temperature-dependent evolutions in PC signals with geometric confinement. The experimental results obtained here not only advance our understanding of fundamental processes in charge reparation but also demonstrated a unique strategy for tailoring the long-range PC response through the SR framework, which may be potentially applied for the design of optoelectronic devices with new functionalities.

## 4. Experimental Section

*Device Fabrication*: The quasi-two-dimensional electron gas (q-2DEG) layer was fabricated through *δ*-doping of silicon substrates, with detailed growth parameters described in Figure S1 (Supporting Information). GaAs nano-constriction devices were patterned using electron beam lithography (EBL) followed by a two-step etching process: The pristine q-2DEG channel was first defined by wet mesa etching, followed by selective argon ion beam etching (IBE) to achieve the final geometrical confined structures. This nanofabrication protocol follows established methodologies similar to those reported in Ref. [27-29].

*SPCM*: Scanning photocurrent measurement (SPCM) was performed using a commercial confocal microscope system (Witec Alpha 300 RAS) equipped with a piezoelectric nano-positioning stage (5-nm resolution). Above-bandgap excitation was provided by a continuous-wave 532 nm laser (2.33 eV photon energy) focused through either: (1) a 100× objective (NA = 0.75, WD = 4.1 mm) for 2D spatial maps (Figures 1b-d, Figure 2a and the insets in Figure 2b) and 1D line profiles of spatial resolved PC (Figure 2b); (2) a 50× long-working-distance objective (NA = 0.55, WD = 9.1 mm) for cryogenic experiments (Figure 4). The laser spot diameter was calculated as $d \approx 1.22\lambda/NA$ for both objectives, yielding values of approximately 865 nm (100× objective) and 1180 nm (50× objective). PC signals were acquired in zero-bias configuration through source-drain electrodes using a low-noise transimpedance amplifier ($10^6$ V/A gain). The 2D spatial maps (Figures 1b-d, 2a, 4a) and 1D line profiles (Figure 2b, Figures 4b-d) of PC were obtained through a piezoelectric nano-positioning stage with step resolutions of 500 nm and 200 nm, respectively, where the specific resolution selection corresponded to different experimental requirements. All cryogenic experiments were performed in a high-vacuum chamber ($3 \times 10^{-5}$ Torr) to suppress surface condensation through the elimination of residual gas adsorption. Notably, the vacuum chamber window introduced < 5% transmission loss at 532 nm, as verified through reference measurements.

## Acknowledgements
X.F. and H.X. contributed equally to this work. This work was supported by the National Key Research and Development Program of China (Grant No. 2024YFA1409800), National Natural Science Foundation of China (Grant Nos. 52403302, 12027805, 12474042, 11991060 and U21A20494), Innovation Program for Quantum Science and Technology (Grant No. 2024ZD0300103), Shanghai Science and Technology Committee (23DZ2260100), Sino-German Center for Research Promotion (Grant No. M-0174) and the Key Project of Westlake Institute for Optoelectronics (Grant No. 2023GD005). H.X. acknowledges Research Professor Wei Yan for the valuable discussions.

## Conflict of Interest
The authors declare no competing financial interest.

## Keywords
bipolar photoresponse, Shockley-Ramo theorem, intervalley scattering, hot electrons

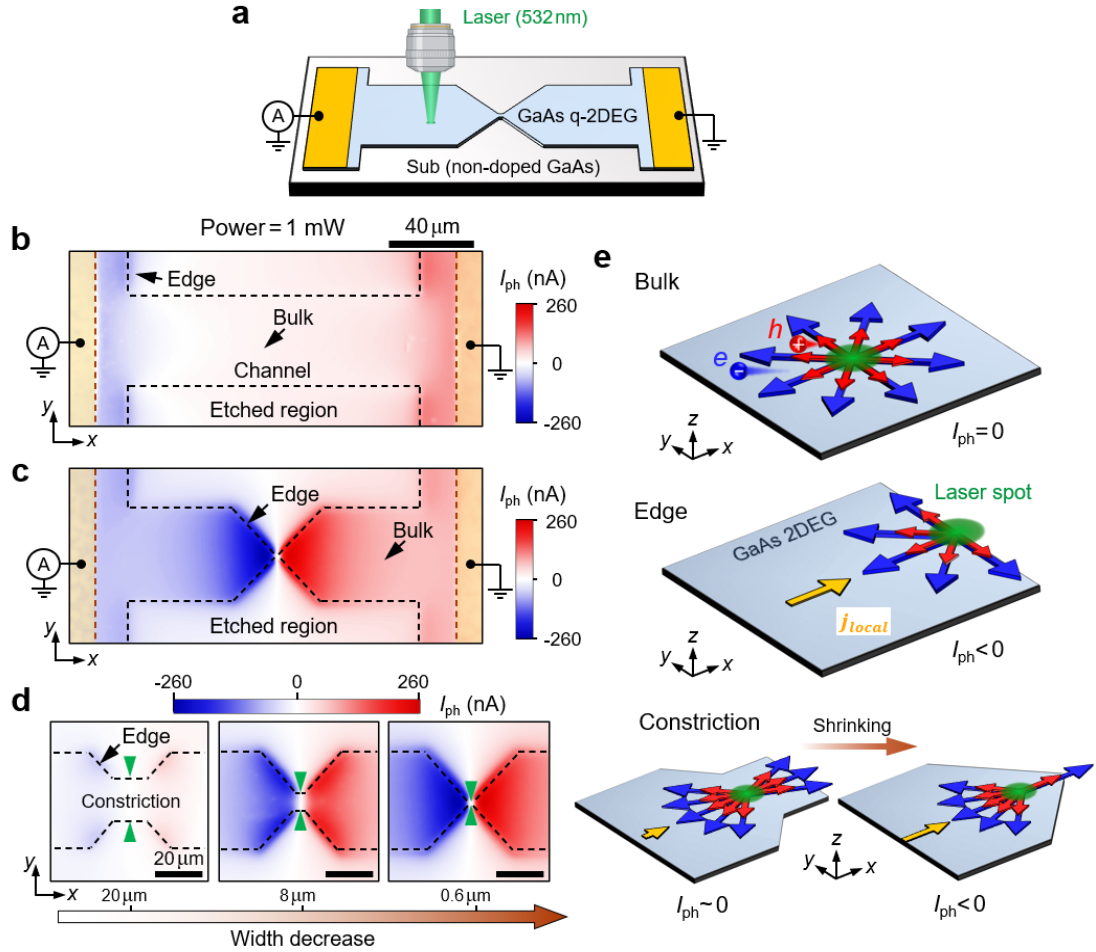

**Figure 1 | Geometrical confinement induced long-range photoresponse.** a) Schematic illustration of SPCM measurement for GaAs-based q-2DEG devices. b) Large-scale SPCM image of a pristine q-2DEG channel without geometrically patterned. c) Large-scale SPCM image of a geometrical patterned q-2DEG channel with a 0.6-µm-width constriction at the middle position. d) Representative SPCM images (scan area: 65 µm × 65 µm) of the constriction regions with widths of 20 µm, 8 µm, and 0.6 µm, respectively as indicated by the green arrows. All the constrictions are fabricated on the middle of the channels. The black dashed lines in (a-c) sketch the contours of the channels. The transparent-yellow regions in (a,b) represent the electrodes, where the edges marked by the red dashed lines. The excitation laser (1 mW) was focused using a 100× objective lens and the focused spot size was about 0.86 µm (see Experimental Section). e) Scheme illustration of local-geometry-induced PC response in bulk, edge, and constriction aeras at 300 K, corresponding to the annotated areas in (b-d). Injecting the bulk areas (top panel), the isotropic diffusion of electrons (blue arrows) and holes (red arrows) induce zero net $j_{local}$ due to the spatial circular symmetry; near the edge areas (middle panel), symmetry-breaking mechanisms—including electron-hole mobility contrast ($\mu_e > \mu_h$) and geometric confinement—generate a net $j_{local}$ (yellow arrow) directed toward channel boundaries; constriction (bottom panel) supply an additional spatial circular asymmetry mechanism to block the carrier diffusion toward the bottleneck, creating unidirectional $j_{local}$, which can be further enhanced by shrinking the constriction width.

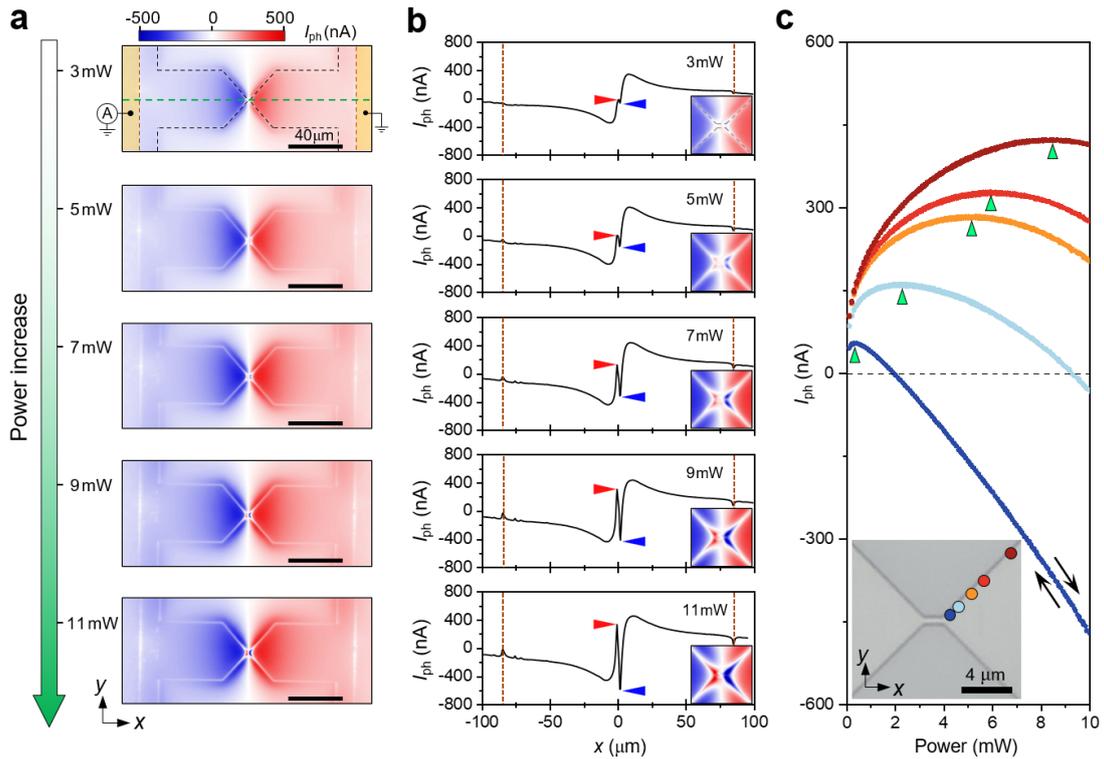

**Figure 2 | Power-dependent anomalous PC behaviors.** a) Large-scale SPCM images (scan area: 200 μm × 85 μm) of q-2DEG channel with 0.6-μm-width constriction under laser power varying from 3 to 11 mW. b) Line profiles of PC along the green dashed line shown in (a) with the corresponding powers in (a). The red dashed lines denote the boundaries of the electrodes. The anomalous PC signals are highlighted by the red/blue triangles that represent the maximum/minimum values of the anomalous PC signals. Inset: representative 6 μm × 6 μm SPCM close-up of the constriction region. c) PC versus laser power curves measured by power cycling testing (0→10→0 mW) as indicated by the black bidirectional arrows taken at multiple coordinates along the constriction edge, as marked by dots in the optical image (inset) with corresponding colors. The green triangles denote power-saturation thresholds exhibiting distance dependence.

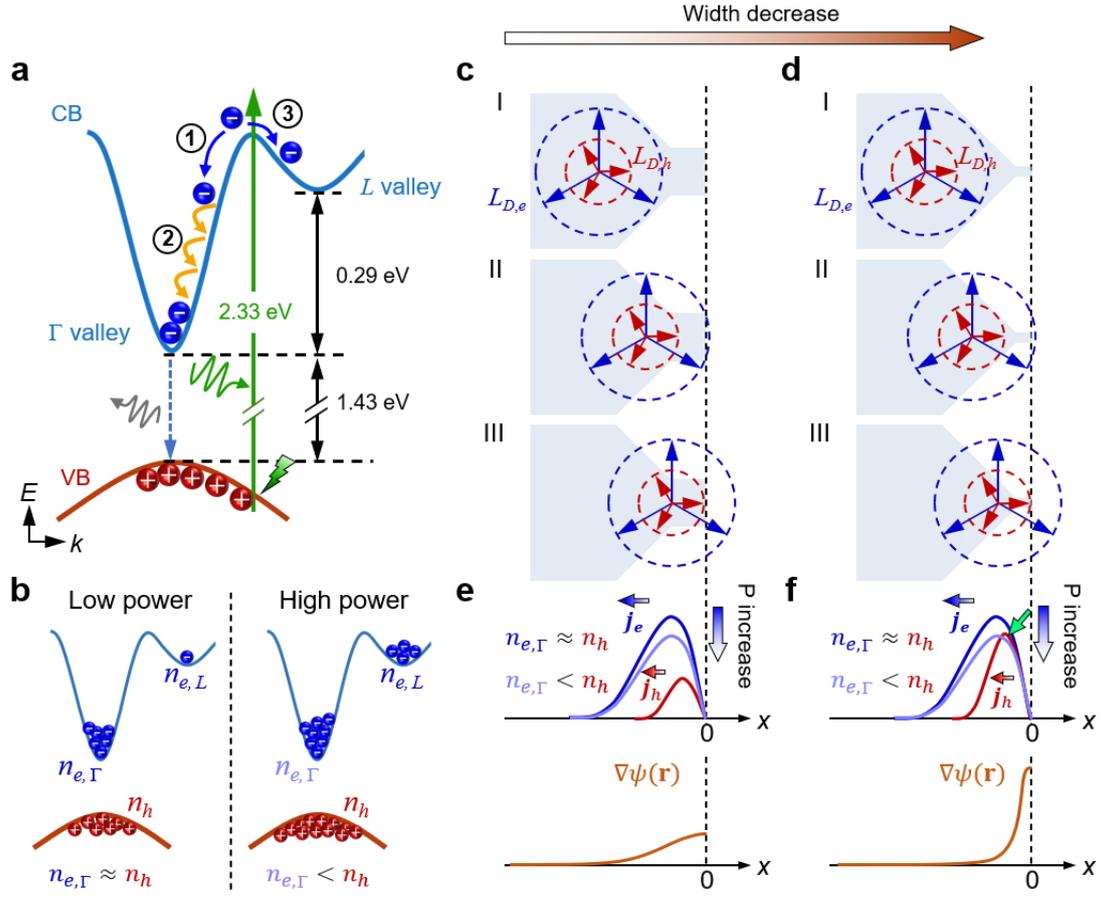

**Figure 3 | The toy model for analyzing the geometry- and excitation-dependent PC signals.**
a) Schematic diagram of the hot-electron intervalley transfer under excitation condition. b) This model based on an assumption: at low power the population of $n_{e,\Gamma}$ is almost equal to $n_h$ (left panel), whereas under high power condition, hot LO-phonon bottleneck effect facilitate the intervalley scattering, causing $n_{e,\Gamma} < n_h$ (right panel). c,d) Illustrations of the geometrical confinement on electron (Γ-valley) and hole diffusions through broad- (c) and narrow- (d) width constrictions at three typical excitation positions on left side of constrictions. The blue and red dashed circles demarcate diffusion boundaries of Γ-valley electrons and holes respectively, with the radii corresponding to their characteristic diffusion lengths $L_{D,e}$ and $L_{D,h}$. e,f) Schematic diagram of 1D distributions of net Γ-valley electron ($j_e$, blue lines) and hole ($j_h$, red lines) diffusion current along the channel, corresponding to the geometrical confinement mechanisms in (c,d). As laser power increases, the ratio of $n_{e,\Gamma}/n_h$ decreases, hence lowering the contribution of $j_e$, as indicated by the suppressed light blue curves in (e,f), and consequently induces an overlap between $j_e$ and $j_h$ in narrow constriction, as indicated by the green arrow in (f). e,f) Schematic 1D distributions of auxiliary weighing field $\nabla\psi$ in broad- (e) and narrow- (f) width constrictions.

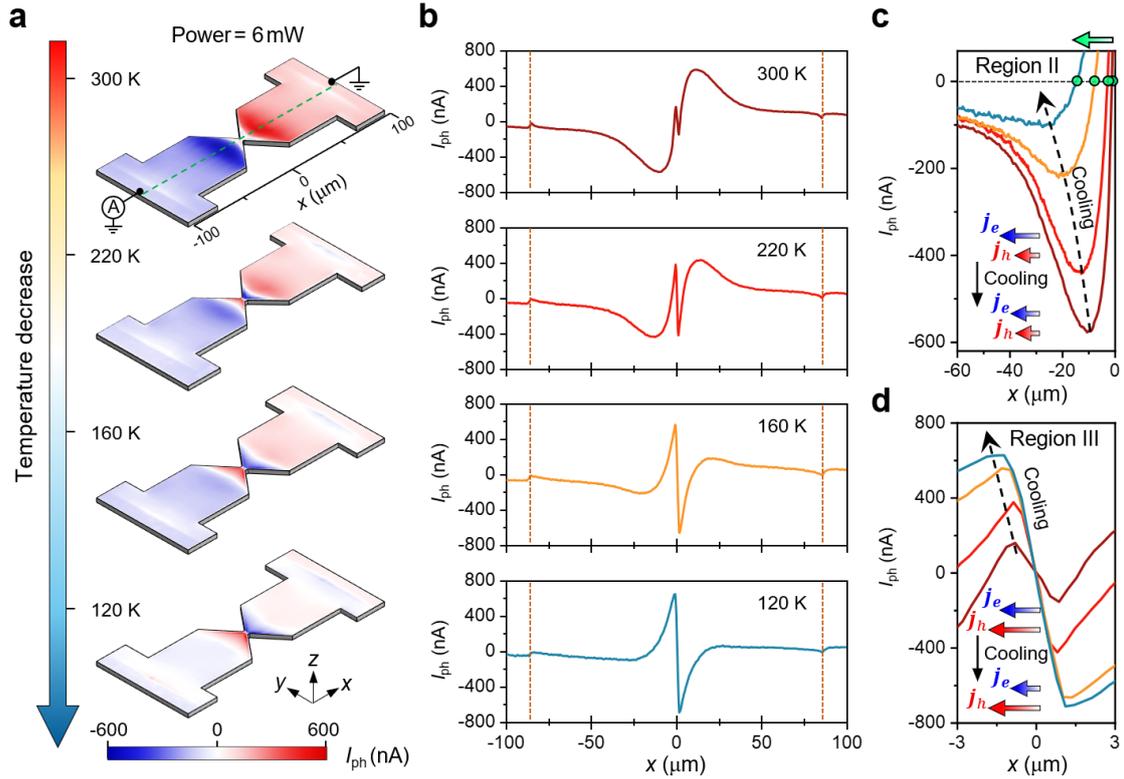

**Figure 4 | Temperature dependence of PC behaviors.** a) Large-area scanning photocurrent microscopy (SPCM) images of a q-2DEG channel containing a 0.6-μm-wide constriction, acquired at cryogenic temperature (300, 220, 160, and 120 K) under a fixed 6 mW laser excitation. The scanned area of 200 μm × 90 μm covers both the constriction region and adjacent electrodes. b) Temperature evolution of PC line profiles extracted along the green dashed trajectory marked in panel (a). The vertical red dashed lines denote the electrode boundaries. c,d) The close-up 1D PC distributions within region II and region III, respectively extracted from (b). The green dots in (c) highlight the crossover points of $j_e$ and $j_h$ (where $j_{local} = 0$, as discussed in Figure 3f) under cooling process. With temperature decreasing, the crossover point shift toward the left direction (indicated by the green arrow). Insets: schematic diagrams of temperature evolution in $j_e$ and $j_h$ on the left side of the constriction. During the cooling process, the intensity of $j_e$ decreases while $j_h$ maintains nearly constant, therefore reducing the $j_{local}$ in region II (c) and enlarging the $j_{local}$ in region III (d).

# Supporting Information

# Geometrical Tailoring of Shockley-Ramo Bipolar Photocurrent in Self-powered GaAs Nanodevices

Xiaoguo Fang[1,2,#], Huanyi Xue[2,3,4,#,*], Xuhui Mao[3], Feilin Chen[2], Ludi Qin[3], Haiyue Pei[2,5], Zhong Chen[6], Pingping Chen[7,8], Ding Zhao[2,4,5], Zhenghua An[3,*], Min Qiu[2,4,5,*]

1. College of Information Science and Electronic Engineering, Zhejiang University, Hangzhou, Zhejiang 310027, China.
2. Zhejiang Key Laboratory of 3D Micro/Nano Fabrication and Characterization, Department of Electronic and Information Engineering, School of Engineering, Westlake University, Hangzhou, Zhejiang 310030, China.
3. State Key Laboratory of Surface Physics and Institute for Nanoelectronic Devices and Quantum Computing, Department of Physics, Fudan University, Shanghai, 200438, China.
4. Westlake Institute for Optoelectronics, Fuyang, Hangzhou, Zhejiang 311421, China.
5. Institute of Advanced Technology, Westlake Institute for Advanced Study, 18 Shilongshan Road, Hangzhou 310024, Zhejiang Province, China.
6. Instrumentation and Service Center for Molecular Sciences, Westlake University, 600 Dunyu Road, Hangzhou 310030, Zhejiang Province, China.
7. State Key Laboratory of Infrared Physics, Shanghai Institute of Technical Physics, Chinese Academy of Sciences, Shanghai 200083, China.
8. University of Chinese Academy of Sciences Beijing 101408, China.

[#] These authors contributed equally to this work.
**\* Authors to whom correspondence should be addressed:** xuehuanyi@wioe.westlake.edu.cn; anzhenghua@fudan.edu.cn; qiu_lab@westlake.edu.cn.

**This file includes:**

**Supplementary Figures S1 – S11**
**Supplementary Notes S1 – S3**
**Supplementary References: S1 – S21**

**Supplementary Figures**

| | | |
|---|---|---|
| Non-doped GaAs | 5 nm | |
| Non-doped Al$_x$Ga$_{1-x}$As (x~0.3) | 8 nm | |
| Non-doped GaAs | 7 nm | |
| Si  2x10$^{12}$/cm$^2$ | | delta-doping |
| Non-doped GaAs | 7 nm | |
| Si  2x10$^{12}$/cm$^2$ | | delta-doping |
| Non-doped GaAs | 7 nm | |
| Si  2x10$^{12}$/cm$^2$ | | delta-doping |
| Non-doped GaAs | 7 nm | |
| Si  2x10$^{12}$/cm$^2$ | | delta-doping |
| Non-doped Al$_x$Ga$_{1-x}$As (x~0.3) | 100 nm | |
| Non-doped GaAs | 500 nm | buffer layer |
| Semi-insulating GaAs substrate | | |

(The middle section containing the four delta-doped GaAs layers is labeled QW.)

**Figure S1. GaAs/Al$_x$Ga$_{1-x}$As quantum well (QW) structure.** The GaAs/Al$_x$Ga$_{1-x}$As heterostructure used in this work is similar to that described in Suppl. Refs. [1-3], grown by molecular beam epitaxy (MBE) on the (100) plane. A quasi-two-dimensional electron gas (q-2DEG) layer with a density $n \approx 1 \times 10^{18}$ cm$^{-3}$ (corresponding to the Fermi energy $E_F \approx 100$ meV at absolute zero temperature $T = 0$ K) and mobility $\mu \approx 1670$ cm$^2 \cdot$V$^{-1} \cdot$s$^{-1}$ [1-3] resides in a 35-nm-thick GaAs QW located 13 nm below the surface.

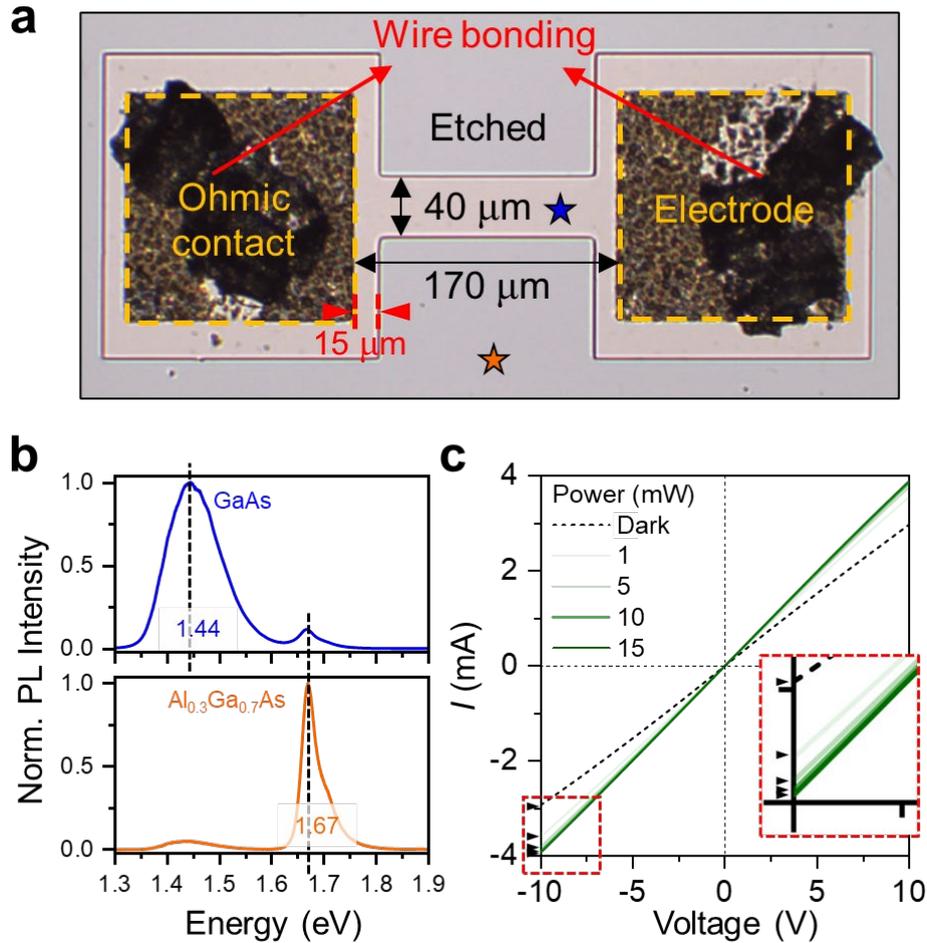

**Figure S2. Structure and properties of the pristine device. a)** Optical image of pristine q-2DEG channel device with a two-terminal structure. The photocurrent (PC) is obtained through the wire bonding to the AuGe/Ni/Au ohmic contacts [1-3]. As a basic structure, all the devices with geometrical-confinement structures in this work are fabricated based on this pristine channel with the width and length of main channel is 40 μm and 170 μm, respectively. **b)** Photoluminescence (PL) measurements taken on the channel (blue curve) and etched region, as marked by the corresponding color stars in (a). The main PL signals taken on the conductive channel predominantly arise from GaAs (measured peak position: 1.44 eV, $E_g \approx 1.4$ eV [4]), while the PL signals of the etched region mainly dominated by Al$_{0.3}$Ga$_{0.7}$As (measured peak position: 1.67 eV, $E_g \approx 1.7$ eV [5]). PL measurements demonstrate that the conducting q-2DEG outside the defined channel region has been completely removed, ensuring no leakage current occurs. **c)** *I-V* curves measured within the linear region of -10 V – 10 V under local photoexcitation (532 nm, 100×) at various power (1-15 mW). The pronounced enhancement (> 20%) in the differential conductance (slope of the curves, as indicated by the intercepts in the inset) under photoexcitation provides direct evidence of photocarriers injection.

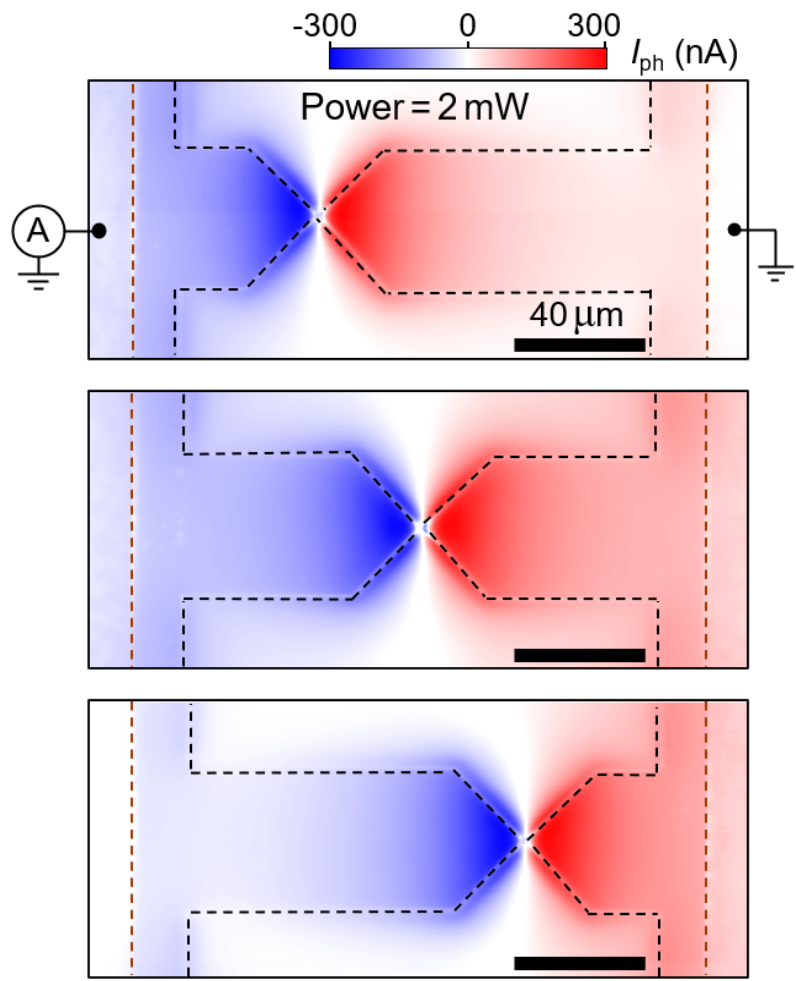

**Figure S3.** Large-scale SPCM images of 0.6-μm-width constrictions asymmetrically/symmetrically positioned at different locations within the channel under low excitation conditions (532 nm, 100 ×, 2 mW). The PC signals exhibit tight clustering on both sides of the constriction, with invariant intensity and polarity that are independent to the relative position of constriction, indicating strong geometry-dependent behavior. The black dashed lines outline the channel boundaries and the vertical brown dashed lines indicate the electrode edges.

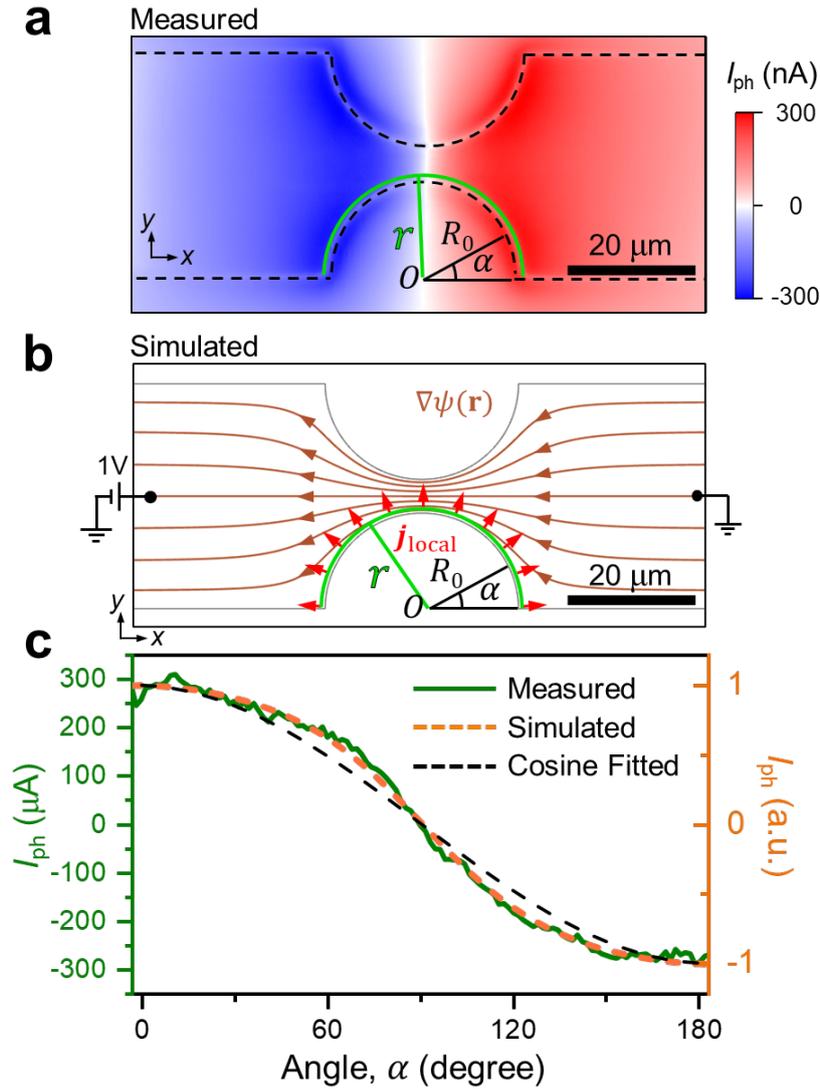

**Figure S4. Preliminary evidence for the Shockley–Ramo-type PC. a)** Measured PC distribution (532 nm, 100 ×, 10 mW) around the geometrically confined device with arc-shape edge structure. The black dashed lines outline the channel boundaries. The width of narrowest region is ~10 μm. **b)** The simulated streamlines (brown lines) of the auxiliary weighting field around the confinement region, where weighting potential at ground contact is zero ($\psi = 0$ V) and at the current collecting contact is set by $\psi = 1$ V [6,7]. Note that the brown arrows denote the direction of $\nabla\psi$, which corresponds to $-\boldsymbol{E}$ [6,7]. The local current $\boldsymbol{j}_{local}$ (red arrows) with normalized intensity perpendicular to the edges, which is primarily governed by electron diffusion (with negative sign). **c)** The measured (green) line profile of global $I_{ph}$ taken along the arc edge (as marked by green arc in (a) with the radius $r = 16$ μm slightly larger than that of arc edges $R_0 = 15$ μm) show the excellent agreement with simulated data (orange dashed line) extracted along the green arc (with radius $r = 16$ μm) in (b). Whereas the measured data cannot be well fitted by cosine function (black dashed line), suggesting that the observed angle dependent PC is not caused by non-intrinsic factors.

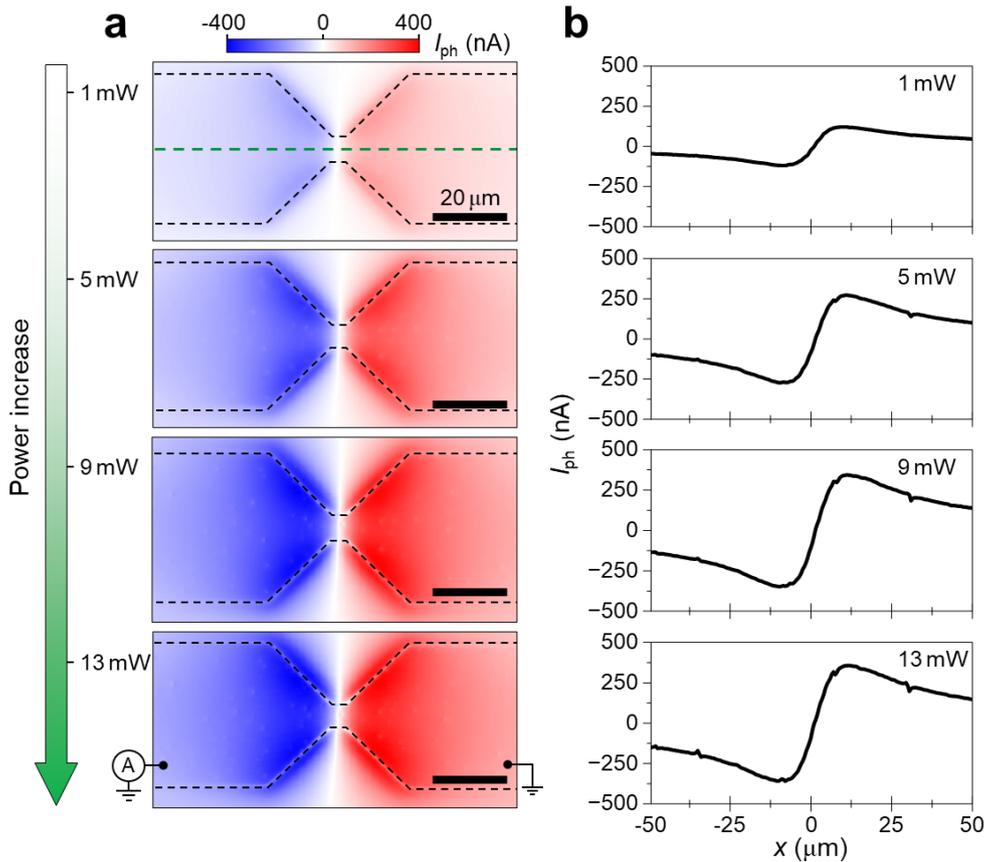

**Figure S5. Power dependence of PC behaviors in an 8-μm-width constriction. a)** PC maps around the constriction region measured at room temperature under various laser powers (1–13 mW). **b)** 1D line profiles of PC extracted along the green dashed line in (a) at corresponding laser powers. In contrast to the PC behaviors observed in the 0.6-μm-width constriction, no anomalous PC signals with reversed polarity are detected in this device even under higher excitation conditions with 13 mW, which exhibit the analogous distributions with fixed profiles. Excitation condition: 532 nm, 100 ×.

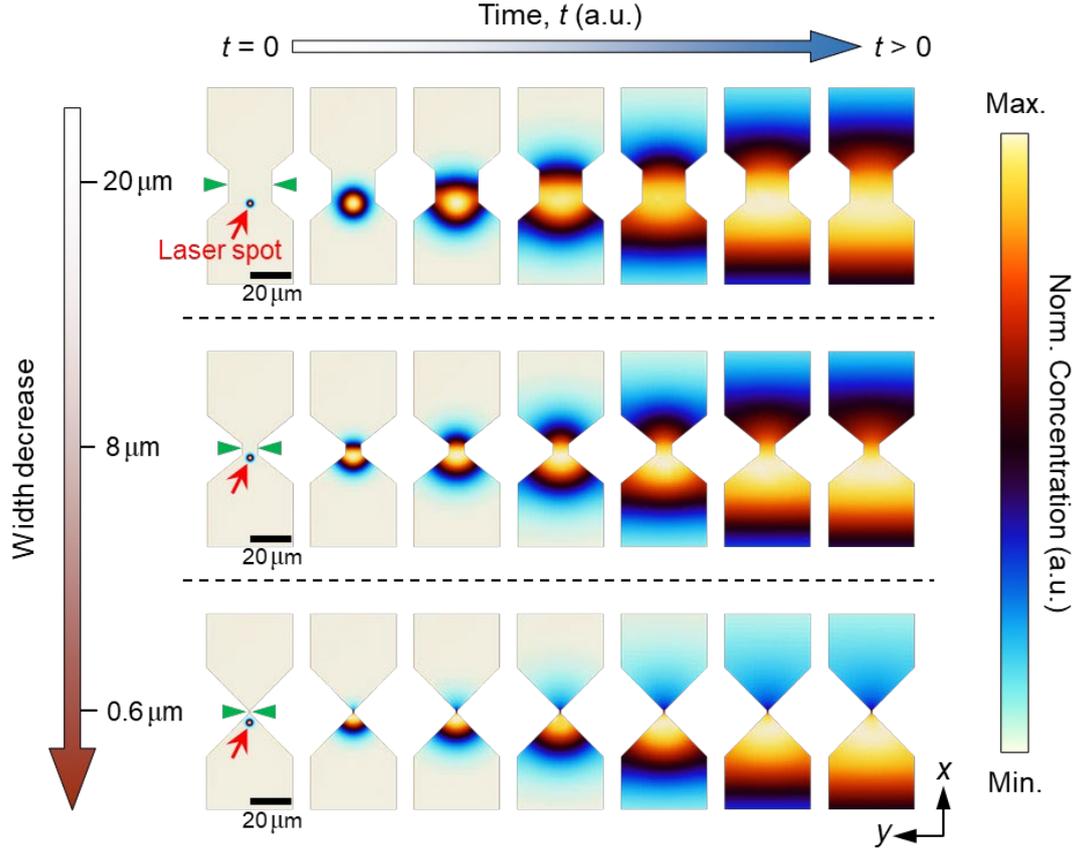

**Figure S6. Simulation of geometric confinement effects on photocarrier diffusion.** We perform finite element analysis to qualitatively investigate the confinement effect of constrictions with various widths (20, 8, and 0.6 μm) governed by the diffusion equation [8]:

$$\frac{dn}{dt} = \nabla \cdot (D\nabla n),$$

where $D$ is the diffusion coefficient (held constant). In this framework, electrons and holes were treated indistinguishably, and their recombination was ignored. At the initial state ($t = 0$), a Gaussian distribution of carriers (full width at half maximum is approximately 1 μm, consistent with the spot size) with a fixed total number ($N$) was positioned at one side of the constrictions to mimic spatially localized laser excitation. According to the above equation, with time evolutions ($t > 0$) the carrier concentration progressively redistributes over time through concentration gradients. The time-dependent evolution clearly demonstrates that photocarriers initially diffuse isotropically into the surrounding medium until they encounter the constriction boundaries. Geometrical confinement subsequently imposes spatial restrictions on carrier redistribution, as evidenced by modified concentration gradients. Notably, progressive narrowing of constriction width (from 20 to 0.6 μm) amplifies this confinement effect, ultimately manifesting as a strikingly asymmetric concentration profile. This width-dependent transition from quasi-bipolar to strongly unidirectional diffusion highlights the critical role of geometrical confinement in tailoring photocarriers dynamics.

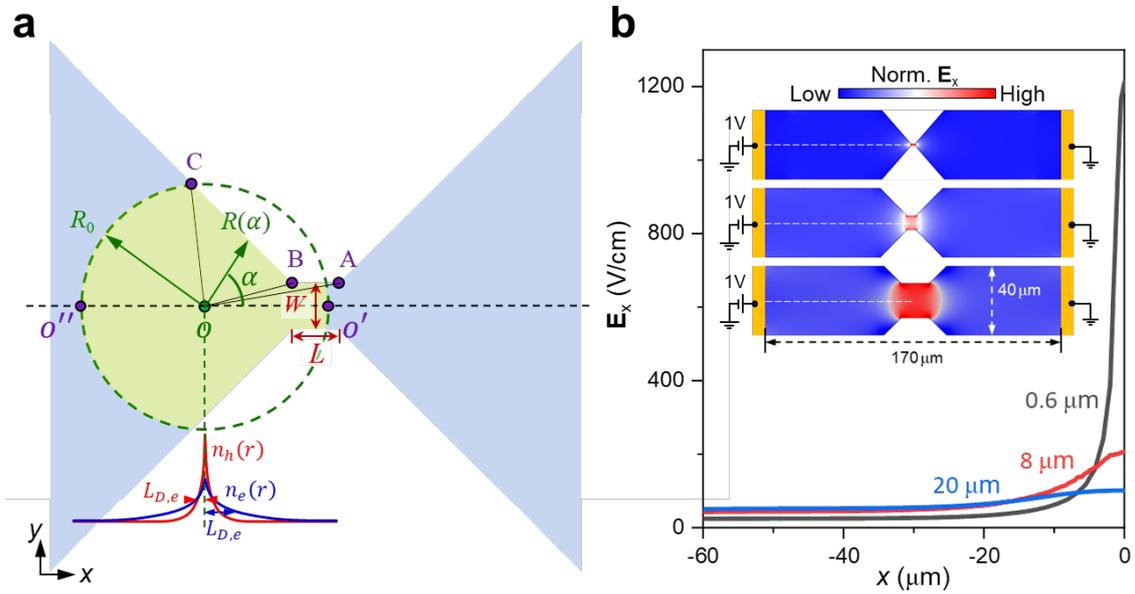

**Figure S7. Schematic of the numerical toy model for simulating the 1D PC distribution under geometrical confinement. a)** The green dashed circle (radius $R_0$) defines the integration limit (Eq. S1) for calculating $j_{e/h}$, where $R_0 \gg L_{D,e/h}$ ensures calculation precision. As the laser position (green dot) scans toward the constriction region (defined by the width $W$ and length $L$) from the left side, the integration limit at angle $\alpha$ is truncated to $R(\alpha)$, which breaks the spatial circular symmetry of carrier diffusion and forms an anisotropic integration region (light green). Characteristic points (purple dots) are selected to segment the angular integration domain, in response to local geometrical confinement, including the kinks of constriction: A and B, the crossover point: C, and the endpoints: $O'$ (right) and $O''$ (left). To simplify the calculations, $L$ was set equal to $W$. **b)** The simulated distributions of auxiliary weighting field $\nabla \psi$ (namely, -$E_x$) of constrictions with various widths by assuming that $\psi$ = 1 V at current-collecting electrodes (left) and $\psi$ = 0 V at ground electrodes (right) [6,7], as shown in the inset of (b). The 1D distributions of $E_x$ are extracted along the channel (as indicated by the white dashed lines in the inset) within 0.6, 8 and 20 μm-width constrictions.

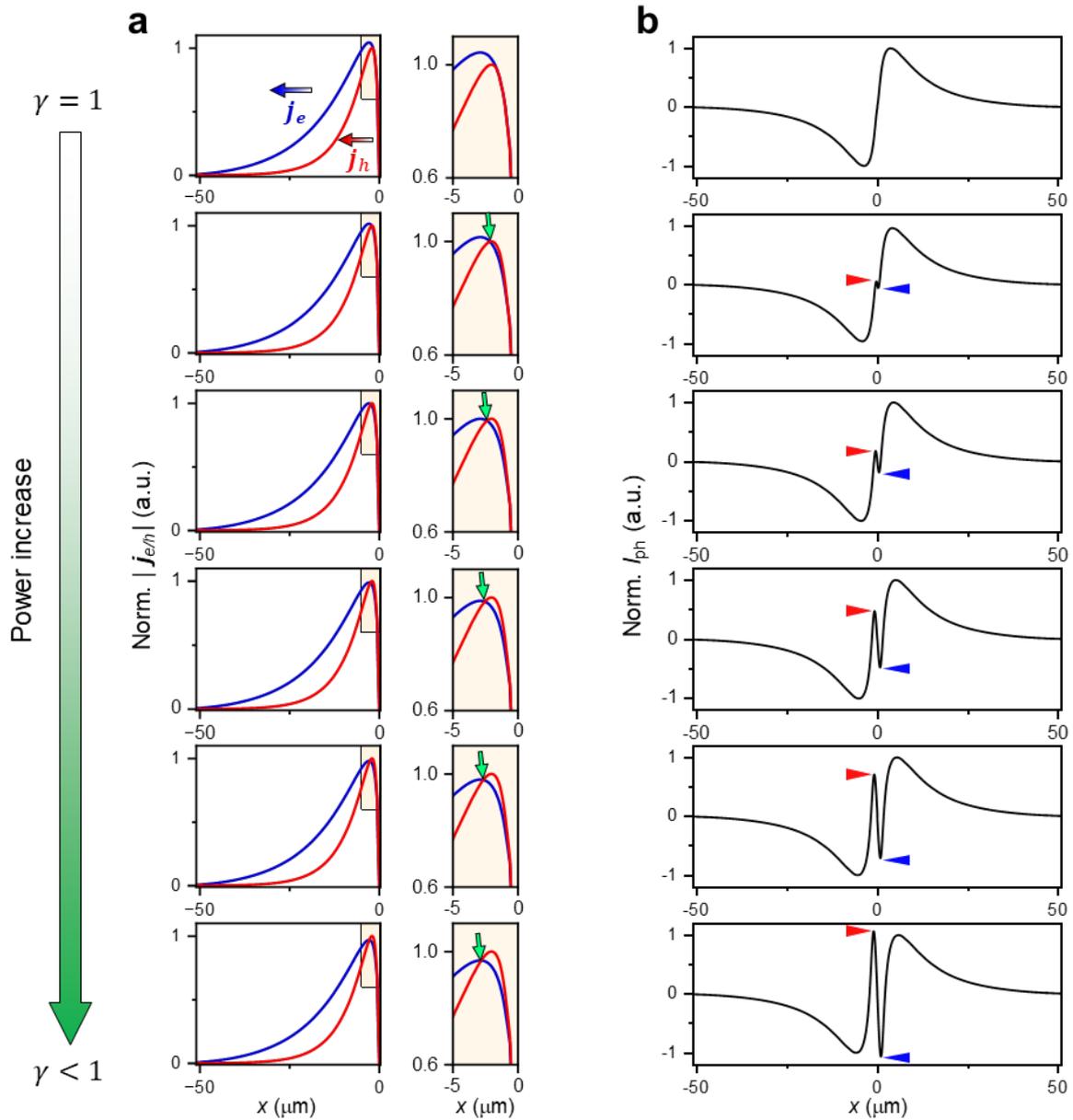

**Figure S8. Reproduction of the power-dependent bipolar PC responses in simulations. a)** Left panel: calculated 1D distributions (*x*: from -50 μm to 0 μm) of $j_e$ (blue lines) and $j_h$ (red lines) at room temperature under increasing power (decreasing factor: $\gamma$ see Suppl. Note. 3), where the intensities are normalized by the peak intensities of $j_h$. Right panel: partial enlargement showing the overlap between $j_e$ and $j_h$, with crossover points highlighted by green arrows. **b)** Line profiles of PC derived from (a) under corresponding laser power conditions. The red/blue arrows mark the peak/deep of anomalous signals. As power increases, the anomalous signals exhibit continuous growth owing to the gradually enhanced overlap between $j_e$ and $j_h$ as shown in (a).

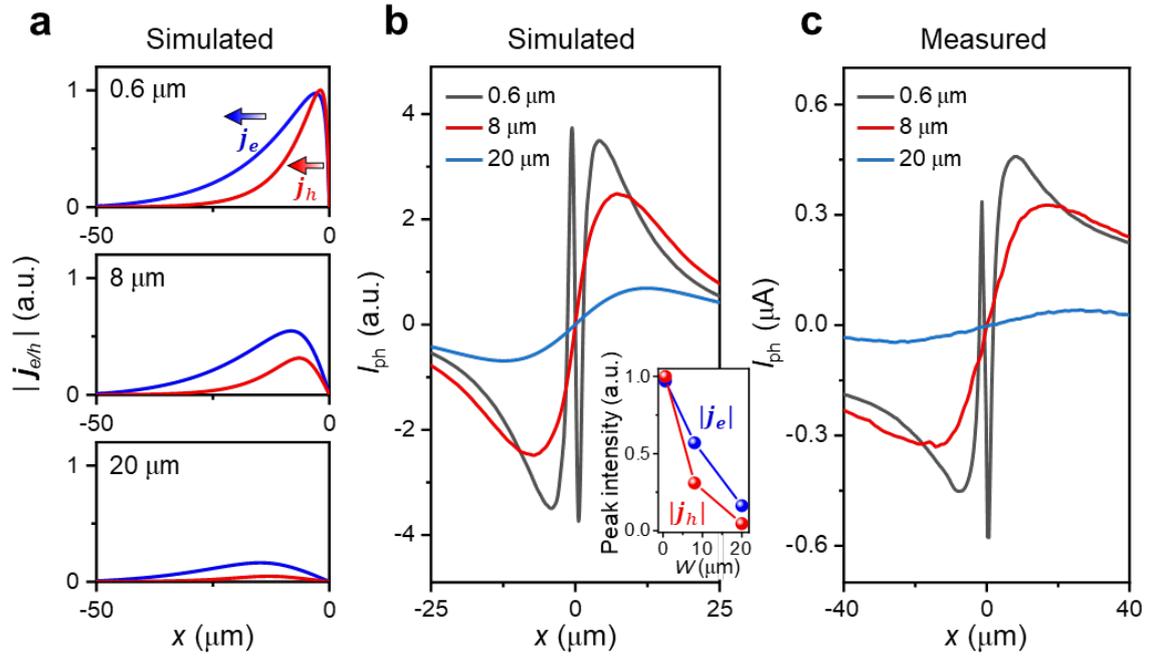

**Figure S9. Geometrical-confinement modulation of PC intensity and line shape with various widths. a)** Simulated 1D distributions (*x*: from -50 μm to 0 μm) of $j_e$ (blue lines) and $j_h$ (red lines) of 0.6, 8, 20-μm-width constrictions at room temperature, where $j_e$ are suppressed by a constant coefficient to mimic 11-mW laser excitation. **b)** Simulated PC line profiles derived from (a) and the respective auxiliary weighting field. Inset: peak intensities of $j_e$ and $j_h$ extracted from (a). **c)** Measured PC line profiles from various constrictions with 0.6, 8, 20 μm widths under 11-mW laser excitation. **A key result in the inset of (b) should be noted**: The intensity of $j_h$ decays faster than that of $j_e$. Consequently, the ratio of peak intensities between $j_h$ and $j_e$ monotonically decreases with increasing constriction width. Therefore, to observe the anomalous signals in wider constriction (i.e., to induce the crossover between $j_e$ and $j_h$ curves), it is required to suppress $j_e$ dramatically with a larger coefficient (i.e., under higher excitation condition). This accounts for the absence of anomalous signals even under high-power conditions (13 mW) in the 8-μm structure, as illustrated in Fig. S5.

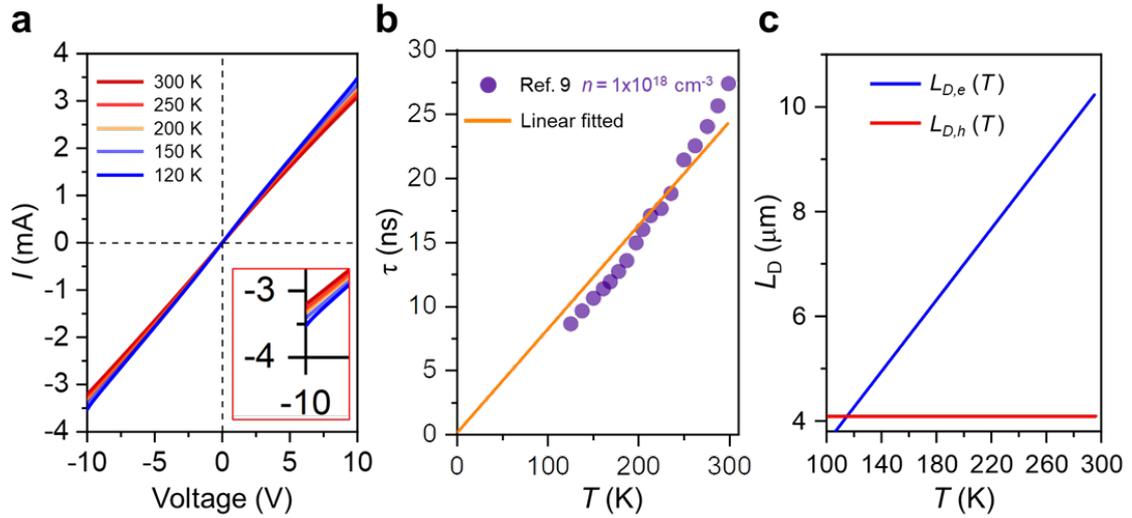

**Figure S10. Temperature dependence of majority (electron) mobility, minority (hole) carrier lifetime and deduced electrons/holes diffusion lengths. a)** *I-V* curves measured in the linear region under dark conditions at cryogenic temperatures show a slight increase in slope ($\propto \sigma$) with decreasing temperature. This trend is consistent with the intercept analysis (inset), where the nearly invariant slopes (intercepts) indicate temperature-insensitive behavior in the majority carrier (electron) mobility, as governed by the relation $\sigma = ne\mu_e$ (where slope $\propto \sigma$, therefore $\mu_e \propto$ slope). **b)** Temperature-dependent minority carrier lifetime $\tau$ (purple dots) extracted from Ref. 9 exhibits an approximately linear trend between 120-300 K, as confirmed by the linear fitting (orange line). **c)** Calculated temperature-dependent diffusion length of electrons $L_{D,e}(T)$ and holes $L_{D,h}(T)$, which deduced by the linear fitted $\tau(T)$ in (b) and respective temperature-dependent mobility $\mu_{e/h}(T)$, see Suppl. Note.2.

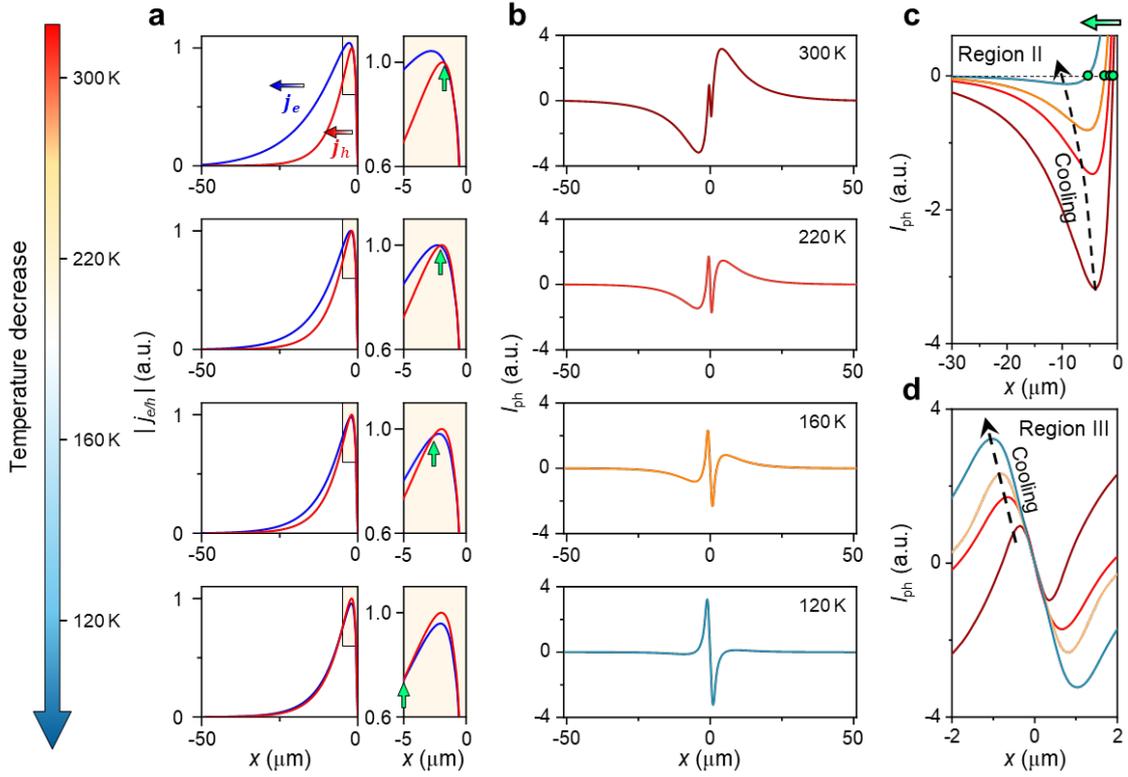

**Figure S11. Reproduction of the temperature-modulated bipolar PC responses in simulations. a)** Left panel: Simulated 1D distributions (*x*: from -50 μm to 0 μm) of $j_e$ (blue lines) and $j_h$ (red lines) at various cryogenic temperatures, where $j_e$ curves are suppressed by the same constant coefficient to mimic the 6-mW laser excitation. Right panel: Enlarged view showing the evolution of overlap between $j_e$ and $j_h$ as temperature decreases, where the crossover points are highlighted with green arrows. **b)** Line profiles of PC derived from (a) at corresponding temperatures, exhibiting trends analogous to experimental data (Fig. 4b in main text). **c,d)** Close-up 1D PC distributions for region II and region III, respectively extracted from (b). The green dots in (c) mark the zero points of PC (where $j_{local} = 0$), which shift leftward as temperature decreases, consistent with experimental results (Fig. 4c in main text). This feature correlates with the temperature-dependent shift of crossover points between $j_e$ and $j_h$, as shown by green arrows in (a).

## Supplementary Notes S1 – S3.

**Supplementary Note S1. Preliminary verification of the Shockley–Ramo-type PC by arc-edge configuration.**

The Shockley-Ramo theorem [6] establishes a fundamental framework for determining the transient photoresponse induced by local mobile charges that far away from the electrodes. Unlike conventional drift-diffusion models, this theorem is based on the fact that the photocurrent detected by the electrode results from the instantaneous change of electrostatic flux lines into or out of the electrodes, rather than from the direct collection of charge carriers per unit time by the electrodes [6].

Recently, the reported SR-type photocurrent (PC) has been achieved by exciting low-symmetry geometries [7,10-13] (e.g., edges and corners), where the local structure breaks the reflection symmetry of carrier transport, generating a net local current $\boldsymbol{j}_{local}$ (requiring electron-hole asymmetry) and distorting the auxiliary weighting fields $\nabla\psi$, consequently inducing a net global PC response.

Inspired by these works, we adopt this widely accepted edge-type [7,10-13] configuration to investigate the SR-type PC. As shown in Fig. S4a and b, the arc-shape edges are featured by a 15-μm radius $R_0$ and 10-μm width at the narrowest region, which allows quantitative analysis of the angle-dependent distribution of PC both in experiment and simulation, thereby enabling the preliminary verification of the SR-type PC in our experiments. In this configuration, the $\boldsymbol{j}_{local}$ is generated dominantly by the electron diffusion current $\boldsymbol{j}_e$ due to the larger diffusion coefficient of electrons $D_e$ compared to that of holes $D_h$ (see Fig. S10), and is normal to boundaries [7,10-13], as sketched in Fig. 1e (in main text). The SPCM image (Fig. S4a) shows that the distribution of PC is concentrated near the arc edges, exhibiting the antisymmetric feature on both sides of the structure with negative sign on the left and positive sign on the right, demonstrating the polarity of PC is dominated by $\boldsymbol{j}_e$.

Following, we adopt a simplified SR theorem (*Song* [6] *and Ma* [7] *et al.*) for calculating the distribution of weighting field $\nabla\psi$, where the weighting potential $\psi$ is set to 1 V at current collecting contact and 0 V at the ground contact, as shown in Fig. S4b. Consequently, $\nabla\psi(x,y) = -\boldsymbol{E}(x,y) = -\left(E_x(x,y)\cdot\boldsymbol{e}_x + E_y(x,y)\cdot\boldsymbol{e}_y\right)$, where $E_x$ and $E_y$ are the x and y components of electric field at position $\boldsymbol{r} = (x,y)$ that can be separately calculated by finite element analysis. Considering the arc-shape edges, we adopt polar coordinates to describe $\boldsymbol{j}_{local}(r,\alpha)$ and $\nabla\psi(r,\alpha)$, as illustrated in Fig. S4a and b, therefore:

$$\nabla\psi(r,\alpha) = -\left(E_x(r,\alpha)\cdot\boldsymbol{e}_x + E_y(r,\alpha)\cdot\boldsymbol{e}_y\right) \tag{S1}$$

In addition, $\boldsymbol{j}_{local}(r,\alpha)$ distributed near the edges with constant radius $r$ exhibits a constant intensity, pointing out to the boundaries [xxx], as indicated by the red arrows in Fig. S4b, hence,

$$\boldsymbol{j}_{local}(r,\alpha) = \text{sign}(e)\cdot\left(\cos(\alpha)\cdot\boldsymbol{e}_x + \sin(\alpha)\cdot\boldsymbol{e}_y\right) \tag{S2}$$

with normalized intensity. Consequently, according to Eq. 2 in main text:

$$I_{ph}(r,\alpha) \approx \boldsymbol{j}_{local}(r,\alpha)\cdot\nabla\psi(r,\alpha) = \cos(\alpha)\cdot E_x(r,\alpha) + \sin(\alpha)\cdot E_y(r,\alpha) \tag{S3}$$

It should be noted that, the boundary condition: $\boldsymbol{n}\cdot\nabla\psi = 0$ [6,7] ($\boldsymbol{n}$ is normal vector) demonstrates that at boundaries ($r = R_0 = 15$ μm), $\boldsymbol{j}_{local}(r,\alpha) = 0$, therefore, the angle-dependent $I_{ph}(r,\alpha)$ was simulated (by Eq. S3) and extracted (from SPCM data) slightly interior of the boundary with $r = 16$ μm to acquire the net $I_{ph}(r,\alpha)$ as shown in Fig. S4c. The simulated and measured $I_{ph}(r,\alpha)$ ($0 \leq \alpha \leq \pi$) exhibit excellent agreement in their line shapes. However, these curves cannot be well fitted by cosine function, further confirming that the observed angle-dependent behavior of PC arises from the SR-type signature rather than from non-intrinsic factors such as the arc-shape induced trivial features.

## Supplementary Note S2. Estimating the diffusion lengths of electrons and holes.

Diffusion lengths of electrons and holes play a crucial role in geometrical confined structures for tunneling the respective local currents. According to the Eqs.1 and 3 in main text, the diffusion lengths of electrons and holes are governed by their mobility $\mu_i$ and lifetime $\tau$ [14]:

$$L_{D,i}(T) = \sqrt{\frac{k_B T \cdot \mu_i(T) \cdot \tau(T)}{|q|}} \tag{S4}$$

where the subscript $i = e/h$ denotes electron or hole, $|q|$ is the elementary charge, $|q| = 1.6 \times 10^{-19}$ C.

According to the Suppl. Ref. [15], the mobility of majorities $\mu_e$ (electrons in this work) is almost temperature-independent in the measured temperature range, as also demonstrated by the little change in *I-V* measurements at various temperature (see Fig. S10a), hence $\mu_e(T) \equiv \mu_e(300 \text{ K}) = 1670 \text{ cm}^2 \cdot \text{V}^{-1} \cdot \text{s}^{-1}$ [1-3]. In contrast, the mobility of minorities $\mu_h$ (holes), exhibits a pronounced temperature dependence: $\mu_h(T) \propto T^{-\alpha}$ where $1 \leq \alpha \leq 2.4$ [15-17] according to the Suppl. Refs. [15-17]. In our simulations, $\alpha$ was taken to be 2, which falls within the range reported in previous works. Thus, $\mu_h(T) = \mu_h(300 \text{ K}) \cdot (\frac{T}{300 \text{ K}})^{-2}$ where $\mu_h(300 \text{ K})$ was intentionally set to $260 \text{ cm}^2 \cdot \text{V}^{-1} \cdot \text{s}^{-1}$, close to the reported value (within $200 - 300 \text{ cm}^2 \cdot \text{V}^{-1} \cdot \text{s}^{-1}$ at 300 K) [15,18].

The temperature-dependent lifetime $\tau(T)$ extracted from the reported data (with comparable doping concentration of ~$1 \times 10^{18}$ cm$^{-3}$) exhibits an approximately linear dependence with temperature in the range of 120-300 K, as shown in Fig. S10b. Using the linear fitted data (Fig. S10b), $\tau(T) \approx \tau(300 \text{ K}) \cdot \frac{T}{300 \text{ K}}$, where $\tau(300 \text{ K}) \approx 24$ ns extracted from the fitted data (Fig. S10b).

Based on these relations, the temperature-dependent diffusion lengths $L_{D,i}(T)$ for electrons and holes were calculated by Eq. S1:

$$L_{D,i}(T) = \begin{cases} L_{D,e}(T) = \frac{T}{300 \text{ K}} \cdot L_{D,e}(300 \text{ K}) \approx \frac{T}{300 \text{ K}} \cdot 10 \text{ μm} \\ L_{D,h}(T) = L_{D,h}(300 \text{ K}) \approx 4 \text{ μm} \end{cases} \tag{S5}$$

Here, $L_{D,e}(300 \text{ K})$ and $L_{D,h}(300 \text{ K})$ are the diffusion lengths of electrons and holes at 300 K, estimated to be about 10 μm and 4 μm, respectively, (as shown in Fig. S10c) close to the reported values [17,19].

**Supplementary Note S3. Investigating constriction-confinement PC behaviors in simulations.**

**3-1. Initial investigation of geometrical-confinement effect.**

Initially, to **directly visualize** how geometric confinement modulates carrier diffusion through the constrictions, we performed semi-quantitative simulations based on the transient-state concentration diffusion equation [8], as discussed in Fig. S6. At the initial stage ($t = 0$ s), a number of electrically neutral particles (with no distinction between electrons and holes) were positioned on one side of the constriction following a Gaussian distribution. As time progresses ($t > 0$ s), driven by the concentration gradient, the particles begin to diffuse radially. With increasing diffusion distance, the geometric confinement effect restricts the particles' passage through the constriction while allowing unimpeded diffusion toward the opposite side, leading to an asymmetric distribution profile (Fig. S6). This asymmetric behavior becomes more pronounced with a shrinking constriction width, directly exhibiting the effect of geometric confinement. Nevertheless, this semi-quantitative approach neglects the crucial electron-hole recombination mechanisms, thereby precluding quantitative calculations of electron and hole diffusion currents under steady-state conditions (consistent with experimental configurations).

**3-2. Quantitative simulation of geometrical-confinement-induced local current.**

As qualitatively sketched in Fig. 3 of the main text, the net local current $j_{local}$ arises from two factors [6,7]: (1) the geometrical confinement-induced net electron/hole current $j_e/j_h$; (2) $e/h$ asymmetry ($j_e + j_h \neq 0$) is prerequired for generating the net $j_{local}$. Here, we adopted a simplified method to calculate the local current $j_{local} = j_e + j_h$ arising from geometrical confinement. As illustrated in Fig. S7a, when the laser spot scans toward the constriction from the left side along the center line, the geometric boundaries of the constriction truncate the diffusion of electrons and holes, especially when the distance between excitation position (circle center) and the boundaries is smaller than the respective diffusion lengths $L_{D,e/h}$. According to the Fick's law [20], the carrier's 1D diffusion at steady state is given by:

$$\begin{cases} n(x) = A \cdot e^{-\frac{x}{L_D}} \\ \frac{dn(x)}{dx} = -\frac{A}{L_D} \cdot e^{-\frac{x}{L_D}} \\ j_{diff}(x) = -D \cdot \frac{dn(x)}{dx} = D \cdot \frac{A}{L_D} \cdot e^{-\frac{x}{L_D}} \end{cases} \quad (S6)$$

where $n(x)$ is 1D distribution of carrier's concentration with the excitation position at $x = 0$, $L_D$ is diffusion length (equal to $L_D = \sqrt{D\tau}$), which describes how fast the concentration decays. Consequently, the concentration gradient $\frac{dn(x)}{dx}$ induces a local diffusion current $j_{diff}(x)$. Both the concentration and diffusion current exhibit exponential decay from the excitation position. Additionally, considering the radiation diffusion of photocarriers (electrons and holes as indicated in subscript) in our configurations, for simplicity (the rigorous mathematical solution in the cylindrical coordinate is very complex [21]) we redefined the Eq. S6 by additionally dividing it by **$2\pi r$**, as shown below:

$$\begin{cases} n_{e/h}^*(r) = \frac{A_{e/h}^*}{2\pi r} \cdot e^{-\frac{r}{L_{D,e/h}}} \\ \frac{dn_{e/h}^*(r)}{dr} \approx -\frac{A_{e/h}^*}{2\pi r \cdot L_{D,e/h}} \cdot e^{-\frac{r}{L_{D,e/h}}} \\ j_{diff,e/h}(r) = -D_{e/h} \cdot \frac{dn_{e/h}^*(r)}{dr} \approx D_{e/h} \cdot \frac{A_{e/h}^*}{2\pi r \cdot L_{D,e/h}} \cdot e^{-\frac{r}{L_{D,e/h}}} \end{cases} \quad (S7)$$

here $A_{e/h}^* = 1/L_{D,e/h}$ ensuring that the total number (normalized by 1) of photoelectrons is equal to that of photoholes $N_e = N_h$, where $N_{e/h} = \int_0^\infty 2\pi r \cdot n_{e/h}^*(r) \cdot dr = 1$. In principle, as demonstrated in Eq. S7a, the diffusion current $j_{diff,e/h}(r)$ decays infinitely far from excitation position. Hence, we performed a simplification in the calculation by assuming that only $j_{diff,e/h}(r)$ within the range of $R_0$ away from the excitation position (as indicated by the green region with the dashed circle boundary in Fig. S7a) contributes to the net current, where $R_0 = 40$ μm far exceeding $L_{D,e/h}$ (Fig. S10c). Therefore, the contribution exceeding the range can be neglected. As a results, if the excitation position (with the coordinate $(x, 0)$, where $x \leq 0$) approaches the left side toward the constriction, and as $|x| < \sqrt{2}R_0$, the integral domain (marked by the green region) for calculating $j_{e/h}(x)$ is truncated by boundaries, as shown in Fig. S7a. Hence,

$$j_{e/h}(x) = 2 \cdot \text{sign}(e/h) \cdot \int_0^{\pi} -\cos(\alpha) \cdot d\alpha \int_0^{R(\alpha)} j_{diff,e/h}(r) \cdot r \cdot dr \quad (S8)$$

where $R(\alpha)$ represents the angle-dependent integration limit, as will be discussed below. Several characteristic points (purple dots in Fig. S7) were selected to partition the angular integration domain, adapting to local geometrical confinement, including the kinks of constriction: A and B, the crossover point: C, and the endpoints in the angle integration domain: $O'$ (lower limit: 0 rad) and $O''$ (upper limit: π rad). Therefore, the angle integration domain $\alpha \in [0, \pi]$ could be divided into four parts: $(0 \leq \alpha < \angle AOO')$, $(\angle AOO' \leq \alpha < \angle BOO')$, $(\angle BOO' \leq \alpha < \angle COO')$ and $(\angle BOO' \leq \alpha \leq \pi)$, where

$$\begin{cases} \angle AOO' = \arctan\left(\dfrac{W}{-L + 2|x|}\right) \\ \angle BOO' = \arctan\left(\dfrac{W}{L + 2|x|}\right) \\ \angle COO' = \dfrac{3\pi}{4} - \arcsin\left(\dfrac{|x|}{\sqrt{2}R_0}\right) \end{cases} \quad (S9)$$

Correspondingly, the distance $R'(\alpha)$ between the excitation position ($O$) to the geometry edges within each integration domain is:

$$R'(\alpha) = \begin{cases} > R_0 \quad \textcolor{green}{\text{obviously!}} & ,(0 \leq \alpha < \angle AOO') \\ \dfrac{1}{2}W \cdot \sqrt{1+\dfrac{1}{\tan^2\alpha}} & ,(\angle AOO' \leq \alpha < \angle BOO') \\ \dfrac{W-2|x|-L}{2(1+\tan\alpha)} \cdot \sqrt{1+\tan^2\alpha} & ,(\angle BOO' \leq \alpha < \angle COO') \\ \geq R_0 \quad \textcolor{green}{\text{obviously!}} & ,(\angle BOO' \leq \alpha \leq \pi) \end{cases} \quad (S10)$$

Hence, the real truncated radius $R(\alpha)$ is defined by a discriminant function within each integration domain:

$$R(\alpha) = \text{Min}\{R'(\alpha), R_0\} \quad (S11)$$

The 1D distribution of $j_e(x)$ and $j_h(x)$ could be calculated respectively from Eq. S8 by substituting Eqs. S5,7,9-11 into Eq. S8.

Finally, the simulated $I_{ph}(x)$ was obtained according to the SR theorem (Eq. 2 in main text): $I_{ph}(x) \approx j_{\text{local}}(x) \cdot \nabla\psi(x) = -E_x \cdot (j_e(x) + j_h(x))$. Here, the 1D distributions of weighting field $\nabla\psi(x)$ ($-E_x$) of corresponding constriction-width devices were extracted from the finite element analysis (the method has been introduced in Supplementary Note. S1 and Supplementary Refs. 6,7), as shown in Fig. S7b.

### 3-3. Reproducing the experimental power-dependent bipolar PC responses.

In our simulations, the power dependence was implemented through the suppression of the current density $j_e(x)$ using a dimensionless scaling factor $\gamma$ ( $0 \leq \gamma \leq 1$), yielding the modified current $j_e^*(x) = \gamma \cdot j_e(x)$. At ultralow excitation powers, photoelectrons predominantly occupy the $\Gamma$ valley, resulting in $n_{e,\Gamma}/n_h \approx 1$ and thus $\gamma \approx 1$. With increasing laser intensity, intervalley scattering depletes the $\Gamma$-valley occupation ( $n_{e,\Gamma}/n_h < 1$ ), which proportionally diminishes the contribution of photoelectrons to $j_e(x)$ via $\gamma < 1$. As shown in Fig. S8, this parameterization successfully captures the anomalous power-dependent behaviors by systematically varying $\gamma$ to emulate discrete excitation power levels. Hence, through the calibrated factor $\gamma$, we could identify the corresponding experimental laser power in our simulations. Importantly, while both $\gamma$ and $n_{e,\Gamma}/n_h$ exhibit monotonic suppression with rising power, they should not be considered equivalent due to the simplified mathematical treatment implemented in our simulations (such as Eq. S7).

**3-4. Investigating the modulation of constriction width on PC behaviors.**

In our experimental observations, the bipolar PC responses initially appeared under 3-mW laser excitation, exhibiting polarity reversal features in the vicinity of the constriction (Fig. 2, main text). However, this feature was absent in wider constriction (8-μm width), even when excited under higher power (13 mW), as shown in Fig. S5. This striking contrast strongly suggests the underlying complex interplay between constriction width and bipolar PC response. Based on our toy model (as previously introduced), 1D distributions of $j_e(x)$ and $j_h(x)$ were simulated within constrictions within constrictions with widths of 0.6 μm, 8 μm, and 20 μm under 11-mW excitation at room temperature. A calibrated factor $\gamma < 1$ was applied to mimic the 11-mW laser excitation. As shown in Fig. S9a, $j_e(x)$ and $j_h(x)$ crossover in the vicinity of the 0.6-μm constriction leading to a pronounced bipolar feature in PC (Fig. S9b). However, the intensity of $j_h(x)$ exhibits a more rapid decline compared to that of $j_e(x)$ as the constriction width increases (as illustrated in the inset of Fig. S9b). Hence, achieving bipolar PC responses through the overlap between $j_e(x)$ and $j_h(x)$ is not possible due to the limited factor $\gamma$ (Fig. S9a), as demonstrated by the corresponding 1D PC line profiles (Fig. S9b). The above simulated features were in excellent agreement with experimental results (Fig. S9c), further verifying the mechanism of the constriction width in modulating the PC behaviors.

**3-5. Reproducing the temperature-modulated bipolar PC responses**.

The simulations of the temperature-dependent PC behaviors were implemented by determining the diffusion lengths of electrons and holes at corresponding temperature (Supplementary Note S2 and Fig. S10). Correspondingly, temperature-dependent $j_e(x)$ and $j_h(x)$ can be deduced by substituting $L_{D,i}(T)$ (Eq. S5) into Eq. S7, as shown in Fig. S11a. Utilizing this method, we successfully reproduced the temperature-modulated bipolar PC responses that were observed in the experiment (Fig. S11b-d).

# Supplementary References